\documentclass[aps,prl,twocolumn,longbibliography,superscriptaddress,final,amsmath,amssymb]{revtex4-2}
\usepackage{graphicx}
\usepackage{dcolumn}
\usepackage{bm}
\usepackage{hyperref}
\usepackage[usenames,dvipsnames]{xcolor}
\usepackage{mathrsfs}

\definecolor{vibrant}{HTML}{E77500}
\definecolor{muted}{HTML}{994400}

\hypersetup{
    colorlinks,
    linkcolor = vibrant,
    citecolor = muted,
    urlcolor = muted
}

\usepackage{braket}
\usepackage{enumitem}

\usepackage[english]{babel}

\newcommand{\beginsupplement}{%
  \clearpage
  \setcounter{section}{0}
  \setcounter{subsection}{0}
  \setcounter{figure}{0}
  \setcounter{table}{0}
  \setcounter{equation}{0}
  \renewcommand{\thesection}{\Roman{section}}
  \renewcommand{\thesubsection}{\Roman{section}.\arabic{subsection}}
  \renewcommand{\thefigure}{S\arabic{figure}}
  \renewcommand{\thetable}{S\arabic{table}}
  \renewcommand{\theequation}{S\arabic{equation}}
}

\newcommand{\Z}[1]{\sigma_{#1}^\mathrm{z}}

\newcommand{\Plus}[1]{\sigma_{#1}^+}
\newcommand{\Minus}[1]{\sigma_{#1}^-}

\newcommand{\kket}[1]{\left.\left|#1\right\rangle\!\right\rangle}

\newcommand{\sket}[1]{\left|#1\right\rangle\!\rangle}
\newcommand{\sbra}[1]{\left\langle\!\left\langle #1\right|\right.}
\newcommand{\sbraket}[2]{\left\langle\!\left\langle #1 \middle| #2 \right\rangle\!\right\rangle}

\newcommand{\kbraket}[3]{\left\langle\!\left\langle #1 \middle| #2 \middle| #3 \right\rangle\!\right\rangle}

\begin{document}

\title{Synchronized Aharonov-Bohm Motifs via Engineered Dissipation}

\author{Christopher W. W{\"a}chtler}
\email{cwwaechtler@icmm.csic.es}
\affiliation{Instituto de Ciencia de Materiales de Madrid ICMM-CSIC, Madrid 28049, Spain}

\author{Gloria Platero}
\affiliation{Instituto de Ciencia de Materiales de Madrid ICMM-CSIC, Madrid 28049, Spain}

\date{\today}

\begin{abstract}
The interplay between external gauge fields and lattice geometry can induce extreme localization dynamics through complete destructive interference. We show that combining this flux-induced localization with engineered dissipation leads to robust spin synchronization in rotationally symmetric spin geometries, referred to as Aharonov–Bohm motifs, with cyclic symmetries of any order. The synchronized dynamics is independent of initial conditions and features entanglement among spins within each motif. We further demonstrate that multiple motifs can fully synchronize when coupled, which is achieved by applying additional collective dissipation acting on all intra-motif spins. These results reveal a direct connection between flux-induced localization, dissipative engineering, and collective quantum synchronization.
\end{abstract}

\maketitle

\textit{Introduction.}---Gauge field fluxes fundamentally influence the dynamics of quantum particles and give rise to many intriguing  phenomena in condensed matter physics, such as the integer and fractional quantum Hall effects~\cite{ vonklitzingQuantizedHallEffect1986a, tsuiTwoDimensionalMagnetotransportExtreme1982, yoshiokaQuantumHallEffect2002}. 
Recent experimental advances of realizing synthetic gauge fields in engineered quantum platforms, including neutral atoms~\cite{linSyntheticMagneticFields2009, aidelsburgerRealizationHofstadterHamiltonian2013, atalaObservationChiralCurrents2014a, anDirectObservationChiral2017}, superconducting circuits~\cite{roushanChiralGroundstateCurrents2017, neillAccuratelyComputingElectronic2021} and trapped ions~\cite{shapiraQuantumSimulationsInteracting2023, wangRealizingSyntheticDimensions2024}, have opened new possibilities for realizing and probing flux-induced effects in a highly controlled way. A prominent example is Aharonov–Bohm (AB) caging~\cite{vidalAharonovBohmCagesTwoDimensional1998}, which appears in lattice models where gauge potentials impose direction-dependent phases that interfere destructively and thereby convert extended Bloch waves into compact localized states, giving rise to perfectly flat (dispersionless) bands in extended lattices. AB caging occurs in a variety of geometries, including rhombic chains~\cite{vidalInteractionInducedDelocalization2000, creffieldCoherentControlInteracting2010a, mukherjeeObservationLocalizedFlatband2015}, Creutz ladders~\cite{creutzEndStatesLadder1999a, zuritaTopologyInteractionsPhotonic2020a}, and the two-dimensional dice lattices~\cite{vidalAharonovBohmCagesTwoDimensional1998}, all of which exhibit complete flux-induced destructive interference at $\pi$-flux (i.e. half a flux quantum) per plaquette. Since its initial discovery in conducting wires~\cite{abilioMagneticFieldInduced1999, naudAharonovBohmCages2D2001},  AB caging has been realized in a variety of other platforms, including photonic lattices~\cite{mukherjeeExperimentalObservationAharonovBohm2018, kremerSquarerootTopologicalInsulator2020}, ultracold atoms~\cite{liAharonovBohmCagingInverse2022a, chenInteractiondrivenBreakdownAharonov2025b}, and superconducting qubits~\cite{zhangSyntheticMultidimensionalAharonovBohm2025}.

Dissipation and decoherence are traditionally regarded as detrimental to coherence-driven phenomena such as AB caging~\cite{liAharonovBohmCagingInverse2022a}. However, when engineered properly and used strategically, dissipation can also act as a resource, for example to prepare nontrivial states~\cite{harringtonEngineeredDissipationQuantum2022b, verstraeteQuantumComputationQuantumstate2009c,  diehlTopologyDissipationAtomic2011b} or to induce nonequilibrium phases and phase transitions~\cite{diehlQuantumStatesPhases2008c, carolloCriticalBehaviorQuantum2019a, marcuzziAbsorbingStatePhase2016a, gillmanNonequilibriumPhaseTransitions2020a, lesanovskyNonequilibriumAbsorbingState2019a, chertkovCharacterizingNonequilibriumPhase2023b}. A prominent example of the interplay between coherent and dissipative dynamics is quantum synchronization~\cite{LS13,WNB14, WNB15, dutta2019critical, cabotQuantumSynchronizationDimer2019a,  thomasQuantumSynchronizationQuadratically2022, eshaqi-saniSynchronizationQuantumTrajectories2020, delmonteQuantumEffectsSynchronization2023a, shenEnhancingQuantumSynchronization2023, xuSynchronizationTwoEnsembles2014a, Waechtler23, zhaoQuantumSynchronizationPerturbed2025a}. Over the past decade, this field has progressed rapidly, with the first experimental realizations probing how classical nonlinear behavior manifests in the quantum regime~\cite{laskarObservationQuantumPhase2020c, koppenhoferQuantumSynchronizationIBM2020b, liExperimentalRealizationSynchronization2025b, taoNoiseinducedQuantumSynchronization2025, Liu25}, and identifying intrinsically quantum aspects of synchronization, in particular its connection to entanglement~\cite{lee2014entanglement, Bruder18_mutual, schmolkeNoiseinducedQuantumSynchronization2022, giorgiQuantumCorrelationsMutual2012a} and quantum interference~\cite{Loerch17, kehrerQuantumSynchronizationInterference2024, kehrerQuantumSynchronizationInterference2024}. More recently, the emergence of synchronized dynamics in condensed-matter and many-body settings has attracted growing attention~\cite{schmolkeNoiseinducedQuantumSynchronization2022, Buca22, Tindall20, loizeauKrylovSpacePerturbation2026a, cakmak2026synchronizationdissipativequantummanybody}, giving rise to concepts such as topological quantum synchronization~\cite{Waechtler23, Waechtler24, liuQuantumSynchronizationOnedimensional2025d}.

In this Letter, we identify a mechanism by which synthetic gauge fields and engineered dissipation cooperate to produce synchronized spin dynamics in two-dimensional spin motifs. Specifically, we show that local dissipation  in rotationally symmetric motifs with $\mathscr C_n$-symmetry (with $n\geq 2$ the order of rotation) induces synchronization of the  magnetization (i.e. $\langle \sigma^z(t)\rangle$), with dynamics reminiscent of AB caging but notably more robust: almost all initial conditions evolve into synchronized motion in stark contrast to the dynamics without dissipation. {Crucially, this synchronization is stable rather than metastable~\cite{Buca22}: it persists in the long-time limit, in contrast to previously identified spin-synchronization mechanisms in which synchronized behavior emerges only as a transient feature during relaxation toward a time-independent steady state~\cite{cabotQuantumSynchronizationDimer2019a}.} We prove that this mechanism is universal to all $\mathscr C_n$-symmetric geometries, which we term Aharonov–Bohm motifs, and derive analytic solutions describing both the dissipative dynamics and the evolution of entanglement, expressed through the concurrence. Furthermore, we demonstrate that synchronization extends beyond individual motifs: inter-motif couplings and collective dissipation channels can fully align the dynamics across interacting AB motifs. Our results have twofold implications: on the condensed-matter side, they establish engineered dissipation as a means to realize correlated dynamics with potential future applications in interference-based flat-band systems~\cite{zuritaTopologyInteractionsPhotonic2020a, zuritaTunableZeroModes2021a, zuritaHiddenTopologyFlatband2025}; on the synchronization side, they reveal a previously unrecognized connection between synchronization and (synthetic) gauge fields, which opens up new avenues for exploring collective quantum behavior.

\begin{figure*}[t]
    \centering
    \includegraphics[width=\linewidth]{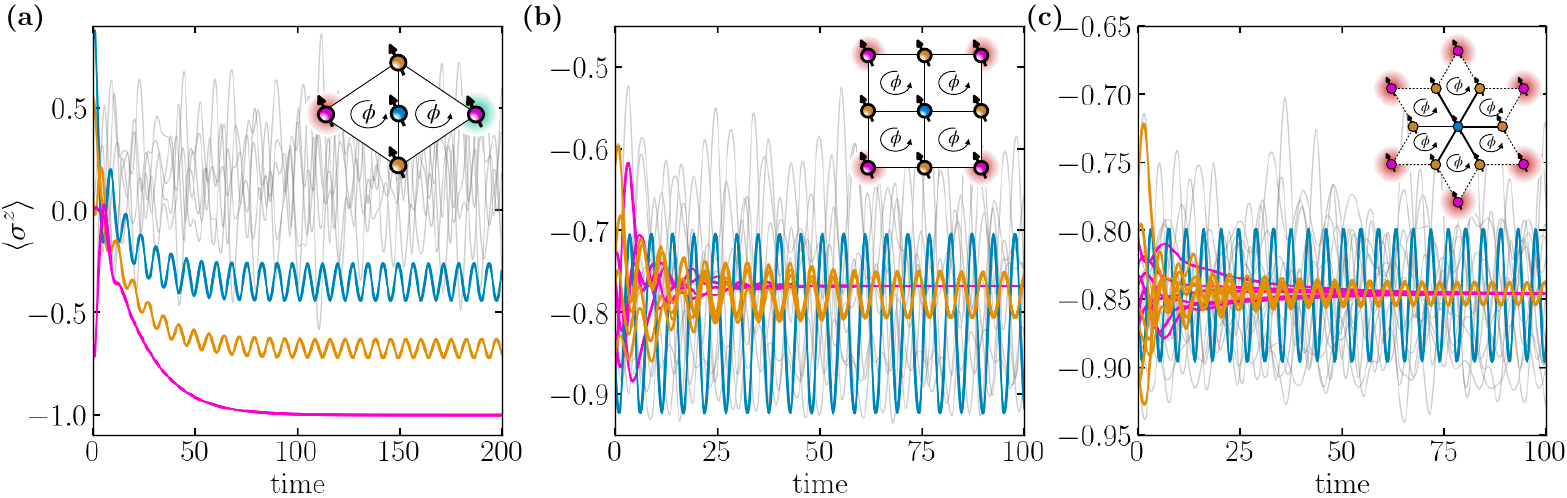}
    \caption{Synchronized dynamics in spin motifs of different cyclic order $\mathscr C_n$ arising from the cooperative interplay between AB interference and locally applied dissipation.
    (a) Motif of order $n =2$  with $N=5$ spins{, where two different dissipators ($\sqrt{\gamma_z}\Z{1, 2}$ (on the left) indicated by a red disk and $\sqrt{\gamma_-}\sigma^-_{2,1}$ (on the right) indicated by a green disk) act on the outer spins}, (b) motif of order $n=4$ with $N=9$ spins {and $\sqrt{\gamma_z}\Z{i, i+1}$ as dissipators}, and (c) motif of order $n=6$ ($N=13$ spins) {with different spin couplings $g_\mathrm{inn} \neq g_\mathrm{out}$ (solid and dashed lines) and $\sqrt{\gamma_z}\Z{i, i+1}$ as dissipators}. Each motif consists of $n$ plaquettes corresponding to the cyclic order $\mathscr C_n$. When a gauge flux $\phi = \pi$ threads each plaquette and dissipation acts on the outer (pink) spins (indicated by the red {and green} discs), the local magnetization $\braket{\Z{\alpha}(t)}$ of the inner (brown) spins becomes synchronized with each other and anti-synchronized with the central (blue) spin. The gray background lines show the corresponding dynamics in the absence of dissipation, starting from the same random initial conditions{~\cite{data}}. {In panels (b) and (c), the initial conditions are restricted to the single magnon subspace $m = -N+2$, which is decoupled from the other magnon subspaces as total magnetization is conserved.}
    {Global parameters: $h = 1.0$, $g_\mathrm{inn} = g_\mathrm{out} = 0.3$, $\phi=\pi$, $\gamma_z = 0.2$. Panel specific parameters: (a) $\gamma_- = 0.1$,  (b) $\gamma_- = 0.0$, and (c)  $g_\mathrm{out} = 0.2$, {$\gamma_- = 0.0$}}.} 
    \label{fig:Motifs}
\end{figure*}

\textit{Dissipative Aharonov–Bohm Motifs.}---We consider a system of interacting spins arranged in plaquettes of four spins. Several plaquettes are connected so that the resulting structure has a $\mathscr C_n$-rotational symmetry, where $n$ is the number of plaquettes meeting at a common central spin; examples of motifs with different cyclic order $n$ are shown in the insets of Figs.~\ref{fig:Motifs}a--c. The total number of spins in a $\mathscr C_n$-symmetric motif is therefore $N = 2 n + 1$. 
 A synthetic gauge flux $\phi$ threads the plaquettes of the motifs. Such fluxes in spin systems can be implemented experimentally in various platforms, including superconducting qubits and trapped ions~\cite{fengQuantumSimulationHofstadter2023, liPerfectQuantumState2018, manovitzQuantumSimulationsComplex2020, shapiraQuantumSimulationsInteracting2023}. 
 The coherent dynamics of the $N$ spins is governed by the Hamiltonian
  {
\begin{equation}
\begin{aligned}
H_{n} =& h\left[\Z{c} + \sum\limits_{i=1}^n \left(\Z{i} + \Z{i,i+1} \right) \right] 
+  g_\mathrm{in}  \sum\limits_{i=1}^n \left(  e^{\mathrm i \phi_{i}} \Plus{c}\Minus{i} + \mathrm{h.c.}\right) \\
&+ g_\mathrm{out} \sum\limits_{i=1}^n \left[e^{\mathrm i \phi_{i,i+1}}\left( \Plus{i}\Minus{i,i+1} + \Plus{i,i+1}\Minus{i+1}\right)
+ \mathrm{h.c.}\right],
\label{eq:PlaquetteHamiltonian}
\end{aligned}
\end{equation}}
where $\Z{\alpha}$ are the Pauli-$z$ operators and  {$\sigma^\pm_\alpha = (\sigma^x_\alpha \pm \mathrm i~\sigma^y_\alpha)/2$} are the usual raising and lowering operators. The parameters $h>0$ and  {$g_\mathrm{inn/out}>0$} denote the on-site field strength and the interaction amplitudes, respectively. The indices in Eq.~(\ref{eq:PlaquetteHamiltonian}) are defined as follows: the central spin is labeled $c$; the \emph{inner} spins, each coupled directly to the central spin, are labeled by a single index $i \in \{1,\dots,n\}$; and the \emph{outer} spins, which connect neighboring inner spins and close the plaquettes, are labeled by a double index $(i,i+1)$. A periodic boundary condition is implied for the indices, i.e., $n+1 \equiv 1$. Each plaquette is therefore formed by the four spins $[c,i,(i,i+1),i+1]$. The total flux threading each plaquette is given by $\phi = \sum_\square \phi_{\alpha}$, where the local phases $\phi_\alpha$ (with $\alpha$ a single or double index) are defined up to a gauge freedom. Throughout, we set $\phi =\pi$ unless otherwise specified.

In addition to the coherent dynamics, we introduce local dissipation acting on the outer spins with index $(i,i+1)$ of each motif. The resulting dissipative dynamics is described by a Lindblad master equation for the system density matrix $\dot \varrho= \mathcal L[\varrho]$ with the Louvillian defined as 
\begin{equation}
\begin{aligned}
\mathcal L [\varrho(t)]=& -\mathrm i [H_n, \varrho(t)] \\
& {+ \sum_{i=1}^n \left(\gamma^\mathrm{z}_{i,i+1} \mathcal D[\Z{{i,i+1}}]\varrho + \gamma^-_{i,i+1} \mathcal D[\sigma^{-}_{{i,i+1}}]\varrho  \right)},
\end{aligned}
\label{eq:Lindblad}
\end{equation}
where $\mathcal D[O]\varrho = O\varrho O^\dagger - \tfrac{1}{2}\{O^\dagger O, \varrho\}$ and  {$\gamma^{z}_{i,i+1}$ and $\gamma^{-}_{i,i+1}$ denote dissipation rates}. Such local dissipation may be realized  {for example} via repeatedly measuring (or replacing) ancillary spins~\cite{cattaneoQuantumSimulationDissipative2023, Liu25}.  {In the following we assume that each outer spin is subject to at least one dissipative process, i.e., either $\gamma^{z}_{i,i+1}\neq 0$ or $\gamma^{-}_{i,i+1}\neq 0$ or both. }  

 {In the absence of $\sigma^-$ dissipation ($\gamma^{-}_{i,i+1}\equiv 0$),} the total magnetization $M = \sum_\alpha \Z{\alpha}$ is conserved by the Hamiltonian, $[H_n, M]=0$, and also commutes with all Lindblad operators , i.e., $[\Z{i,i+1}, M]=0$. The dynamics  {thus} decomposes into independent magnetization sectors labeled by $m \in \{-N, -N+2, \dots, N\}$. As shown in the Supplementary Material~\cite{SM}, all sectors except the single-magnon subspaces with $m = \pm(N-2)$ possess a unique steady state, whereas the single-magnon sectors support non-stationary dynamics with the Liouvillian restricted to them having purely imaginary eigenvalues~\cite{Buca19, Buca22}.  {Finite values of $\gamma^{-}_{i,i+1}$, however, introduce additional dissipative processes that break magnetization conservation and establish unidirectional coupling between sectors. These processes drive population toward the fully de-excited state with $m=-N$. Notably, within the single-magnon subspace $m = -N + 2$, there exists a protected subspace that remains decoupled from the dissipative cascade. This protected subspace is precisely the origin of the synchronized dynamics observed in the system; cf. Supplementary Material~\cite{SM}. In the following, we therefore focus our analysis on this protected subspace. Note, that all considerations equally apply to the case where  the system is driven into the fully excited state ($m=N$) by replacing all $\sigma^-_{i,i+1}$ with $\sigma^+_{i,i+1}$ in Eq.~(\ref{eq:Lindblad}), in which case the protected subspace lies within the $m=N-2$ sector.}

In Fig.~\ref{fig:Motifs} we show the time evolution of the local magnetization $\braket{\Z{\alpha}(t)}$ for three different motifs starting from random initial conditions within the single-magnon subspace: the $\mathscr C_2$-symmetric motif in panel (a), the $\mathscr C_4$-symmetric motif in panel (b), and the $\mathscr C_6$-symmetric motif in panel (c). The insets depict the corresponding AB motif, with the central spin shown in blue, the inner spins in brown, and the dissipative outer spins in pink. Despite the different cyclic orders $n$  {and different dissipative processes}, all motifs exhibit the same qualitative dynamics. The outer (pink) spins relax to a stationary magnetization on a timescale set by the dissipation rate $\gamma$, whereas the central (blue) spin and the inner (brown) spins retain oscillations of finite amplitude. All inner spins evolve in full synchrony, while the central spin oscillates in anti-phase with them.

The observed dynamics arises from the interplay between quantum interference and dissipation. In the system without dissipation and in the presence of a gauge flux $\phi =\pi$, tunneling pathways interfere destructively through the AB effect, giving rise to localized wave packets (and flat bands in extended lattices), so-called AB cages~\cite{vidalAharonovBohmCagesTwoDimensional1998}. In the AB motifs considered here, a local excitation of the central spin can coherently populate the inner spins, while destructive interference suppresses excitation of the outer spins. 
This behavior, however, occurs only when the initial state is fully localized at the central spin. For random initial conditions, multiple eigenmodes of the Hamiltonian with different frequencies contribute, resulting in unsynchronized dynamics, as illustrated by the gray lines in the background of Figs.~\ref{fig:Motifs}a--c. When dissipation is introduced, any component of the dynamics not confined to the AB-caged subspace is damped out. The long-time dynamics is therefore restricted to this interference-protected subspace and exhibits oscillations reminiscent of AB caging in extended lattices. For this reason, we refer to these structures as AB motifs.

The dissipative dynamics of $\mathscr C_n$-symmetric motifs can be captured using concepts from quantum synchronization  {and the emergence of oscillating coherences \cite{zhaoQuantumSynchronizationPerturbed2025a}}; cf. Appendix~A. The Liouvillian $\mathcal L$ corresponding to  Eqs.~(\ref{eq:PlaquetteHamiltonian}) and (\ref{eq:Lindblad})  {fulfills the algebraic relations necessary for the existence of a pair of purely imaginary eigenvalues $\mathrm i\Omega_\pm  = \pm \mathrm i 2\sqrt{n} g_\mathrm{inn}$} with corresponding Liouvillian eigenmodes  {$\varrho_{+-}= \ket{\psi_+}\bra{\psi_-}$ and $\varrho_{-+}= \ket{\psi_-}\bra{\psi_+}$} (see the Supplementary Material~\cite{SM} for the proof). The associated states are given by~\cite{SM} 
 {\begin{equation}
\label{eq:PsiPlusMinus}
    \ket{\psi_\pm} = \frac{1}{\sqrt{2}}\left(\pm\ket{c} + \frac{1}{\sqrt{n}}\sum\limits_{i=1}^n \ket{i}\right) 
\end{equation}
}
with $\ket{j}$ denoting the configuration with a single spin excitation at site $j\in\{c, 1,\dots, n\}$, and inherit the rotational $\mathscr C_n$ symmetry of the motif. 
The long-time dynamics within the single-magnon subspace is dominated by the oscillatory contributions of  {$\varrho_{\pm\mp}$}. For generic initial conditions $\varrho(0)$ with nonzero overlap with this  {protected subspace} and after a transient time $\tau$, the local spin magnetization of the central and inner spins for $t \geq \tau$ becomes (up to exponentially small corrections)
\begin{align} 
\label{eq:LocalMag1}
   \braket{\Z{c}(t)} &=  A_c - 2 \left|\braket{\psi_+|\varrho(0)|\psi_-}\right|\cos(\Omega t+\varphi), \\ 
   \braket{\Z{i}(t)}& = \bar A + \frac{2}{n} \left|\braket{\psi_+|\varrho(0)|\psi_-}\right|\cos(\Omega t+\varphi), \label{eq:LocalMag2}
\end{align}
while the outer spins approach a constant value $\braket{\Z{i,i+1}(t)} = \tilde A$. The constants $A_c$, $\bar A$ and $\tilde A$ are determined by the initial conditions and specified in the Supplementary Material~\cite{SM} {; in the case of local spin relaxation $\gamma^-_{i,i+1}\neq 0$, $\tilde A = -1$.} The phase $\varphi$ is defined via $\braket{\psi_+|\varrho(0)|\psi_-} = \left|\braket{\psi_+|\varrho(0)|\psi_-}\right|\exp(\mathrm i \varphi)$. Importantly, the same phase governs all oscillating spins, leading to synchronized motion within the motif as a direct consequence of quantum interference combined with spatially structured dissipation. This mechanism is considerably more robust than the purely coherent case: synchronization emerges for almost all initial states, except those orthogonal to the  {protected subspace}. The oscillation amplitude is maximal when the system is initialized with a single spin flip at the motif center, i.e. $\varrho(0) = \ket{c}\bra{c}$.  {Note that in the case where one would remove the outer (pink) spins and start with random initial conditions in the smaller lattice, all eigenstates participate in the dynamics and thus no synchronization in terms of $\braket{\Z{i}(t)} = \braket{\Z{j}(t)}$ for all $i,j$ and $t\geq \tau$ is observed in general.}

\begin{figure}
    \centering
    \includegraphics[width=\linewidth]{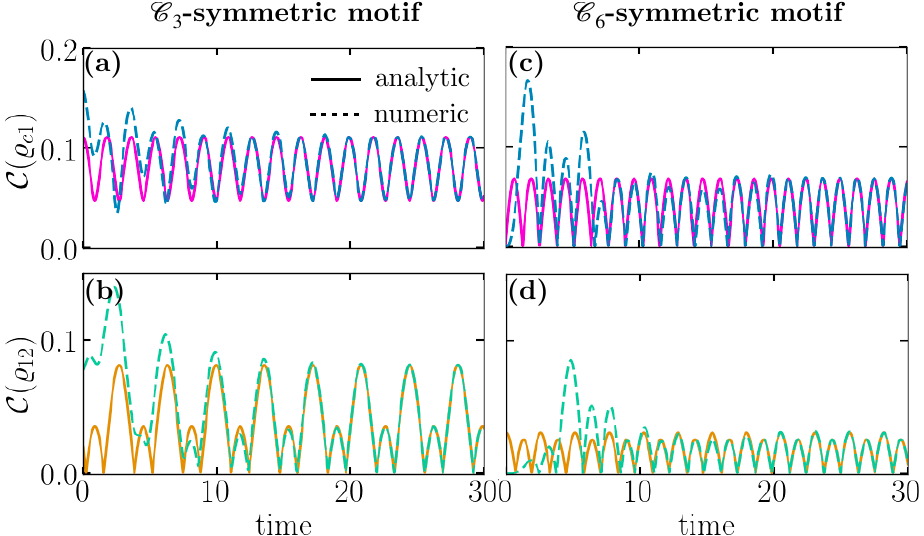}
    \caption{Concurrence as a measure of entanglement during the dissipative dynamics. Panels (a) and (b) show results for the $\mathscr C_3$-symmetric motif with $N=7$ spins, and panels (c) and (d) for the $\mathscr C_6$-symmetric motif with $N=13$ spins. The top panels display the concurrence between the central spin and an inner spin $\mathcal C(\varrho_{ci})$, which is identical for all inner spins due to rotational symmetry. The bottom panels show the concurrence between two inner spins $\mathcal C(\varrho_{ij})$, also identical for all inner-spin pairs. Solid lines correspond to the analytical expressions in Eqs.~(\ref{eq:concurrenceC})–(\ref{eq:concurrence}), including a numerically determined constant shift, while dashed lines show the full numerical results obtained from solving the Lindblad equation and include damped dynamics.  {For panels (a) and (b),  we choose random initial conditions within the single magnon subspace, while for panels (c) and (d) we start with a single excitation on site $i=5$~\cite{data}}. The concurrence exhibits oscillations in time, reflecting the underlying synchronized spin dynamics.  {Parameters: $h = 1.0$, $g_\mathrm{inn} = g_\mathrm{out} = 0.5$, $\phi =\pi$, $\gamma^{z}_{i,i+1} = 0.2$, $\gamma^{-}_{i,i+1} = 0.0$.}}
    \label{fig:Entanglement}
\end{figure}

\textit{Entanglement.}---Interactions between spins in many-body systems naturally generate entanglement during their evolution. In open systems, however, dissipation and dephasing typically suppress such correlations~\cite{diasEntanglementGenerationDistant2023}. Nevertheless, earlier studies have reported connections between synchronization and entanglement generation~\cite{lee2014entanglement,Bruder18_mutual,schmolkeNoiseinducedQuantumSynchronization2022,taoNoiseinducedQuantumSynchronization2025}, which motivates us to examine whether the synchronized spin dynamics in the AB motifs is accompanied by entanglement. To this end, we evaluate the concurrence, defined as $\mathcal C(\varrho_{\alpha\alpha'}) = \max\{0, \lambda_1 - \lambda_2 - \lambda_3 - \lambda_4\}$, where $\varrho_{\alpha\alpha'} = \mathrm{Tr}_{\mathcal A\setminus(\alpha,\alpha')}[\varrho(t)]$ is the reduced density matrix of spins $\alpha$ and $\alpha'$, and $\mathcal A$ denotes the collection of all indices $\alpha$. The quantities $\lambda_k$ are the square roots of the ordered eigenvalues of the matrix $\varrho_{\alpha\alpha'}\tilde{\varrho}_{\alpha\alpha'}$, with $\tilde{\varrho}_{\alpha\alpha'}$ denoting the spin-flipped state~\cite{woottersEntanglementFormationArbitrary1998}. Owing to the rotational symmetry of the motifs, it is sufficient to consider only the concurrence between the central spin and any inner spin, $\mathcal C(\varrho_{ci})$, and that between two inner spins, $\mathcal C(\varrho_{ij})$. Using this symmetry, we derive analytical expressions for the concurrence after the transient time, i.e., for $t \geq \tau$~\cite{SM},
\begin{align}
\label{eq:concurrenceC}
        \mathcal C(\varrho_{ci}) &= 2 C_c +\frac{1}{\sqrt{n}}\left|\tilde C_+ - 2\mathrm i \left|\braket{\psi_+|\varrho(0)|\psi_-}\right|\sin(\Omega t + \varphi)\right|, \\
\mathcal C(\varrho_{ij}) &= 2 \bar C +\frac{1}{{n}}\left|\tilde C_- +2 \left|\braket{\psi_+|\varrho(0)|\psi_-}\right| \cos(\Omega t + \varphi)\right|,   \label{eq:concurrence}    
\end{align}
where $C_c$ and $\bar C$ denote contributions
from components of the density matrix outside the  {protected synchronization} subspace (which we obtain numerically below), and $\tilde C_\pm = \braket{\psi_+|\varrho(0)|\psi_+} \pm \braket{\psi_-|\varrho(0)|\psi_-}$. The oscillation frequency $\Omega$ and phase  $\varphi$ coincide with those in Eqs.~(\ref{eq:LocalMag1}) and~(\ref{eq:LocalMag2}). Note that for the system at hand, owing to the rotational symmetry, the concurrence is bounded by $ 0\leq \mathcal C(\varrho_{ci}) \leq 1/\sqrt{n}$ and $ 0\leq \mathcal C(\varrho_{ij}) \leq 1/n$~\cite{SM}.

Figure~\ref{fig:Entanglement} shows $\mathcal C(\varrho_{ci})$ [panels (a) and (c)] and $\mathcal C(\varrho_{ij})$ [panels (b) and (d)] for the $\mathscr C_3$ and $\mathscr C_6$ motifs  {with $\gamma^-_{i,i+1} \equiv 0$}. The solid lines are the analytical expressions from Eqs.~(\ref{eq:concurrenceC})–(\ref{eq:concurrence}), using numerically determined constants $C_c$ and $\bar C$; the dashed lines show the full numerical solutions of the Lindblad equation. After a short transient $\tau$, the analytical and numerical results coincide. As the system size increases (larger $n$), the concurrence amplitude decreases, consistent with the reduced synchronization amplitude. The maximal concurrence within the  {synchronization} subspace occurs for an initially localized excitation at the central spin, which mirros the behavior of the synchronization amplitude~\cite{SM}.  {The existence of entanglement confirms that the synchronized dynamics is a genuinely quantum phenomenon, rather than merely classically correlated spin flips. }

\textit{Disorder.}---Synchronization in the motifs relies on the exact cancellation of two tunneling paths at the outer spins, which may be difficult to realize perfectly in experiments. We therefore examine how disorder affects the synchronized spin dynamics. We first note that variations in the local dissipation rates  {$\gamma^z_{i,i+1}$ and $\gamma^-_{i,i+1}$} do not alter the long-time synchronized behavior as long as the dissipation channels remain strictly local. We thus focus on perturbations of the coherent dynamics by introducing disorder into $H_n$ of Eq.~(\ref{eq:PlaquetteHamiltonian}), i.e., we consider  $\tilde H_n = H_n + \varepsilon V$, where $V$ can represent  on-site or coupling disorder, including perturbations of the flux per plaquette. 

Using non-Hermitian perturbation theory, one finds explicitly~\cite{SM} that the first-order correction to the oscillatory Liouvillian eigenmodes with purely imaginary eigenvalues is again purely imaginary, producing only a frequency shift. A negative real correction can appear only at second order, leading to gradual damping of the synchronized dynamics and ultimately to a unique stationary state. However, for sufficiently weak perturbations,  {$\varepsilon^2 \ll g_\mathrm{inn/out}, \gamma^{z,-}_{i,i+1}$}, the induced decay is slow: many oscillations occur before damping becomes appreciable, which renders the synchronized dynamics observable for long times. This regime corresponds to meta-stable synchronization~\cite{Buca22}. 

\begin{figure}
    \centering
    \includegraphics[width=\linewidth]{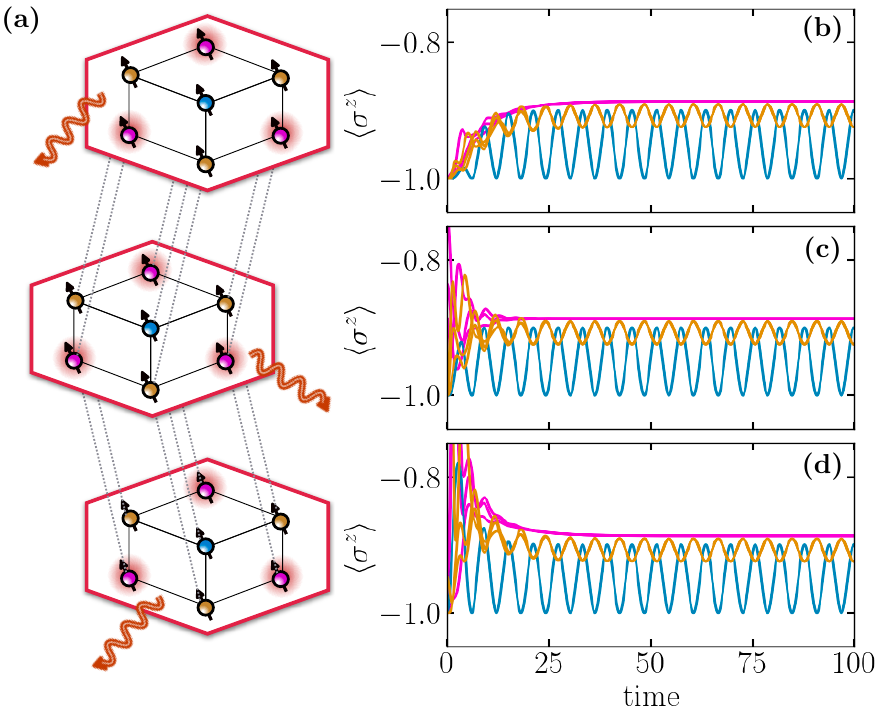}
    \caption{(a) Schematic representation of three coupled $\mathscr C_3$-symmetric motifs. In addition to the intra-motif couplings, each spin in the middle motif is coupled to the corresponding spins in the top and bottom motifs (dashed lines). Collective dissipation, $L_\nu =\sqrt{\kappa} \sum_{\alpha\in \mathcal A} \Z{\alpha\nu}$ indicated by the red outline, acts on all spins within a motif labeld by $\nu$.
Panels (b)–(d) show the magnetization dynamics of the motifs shown on the left of each panel {, starting from random initial conditions within the single-excitation subspace~\cite{data}}. After a short transient, the spins subject to strictly local dissipation (red discs) relax to stationary values (pink lines). The central (blue) spins become fully synchronized, both in amplitude and phase, across all three motifs. The inner (brown) spins also synchronize across motifs, with an oscillation amplitude equal to one third of that of the central spins, and evolve in anti-phase with them. 
 {Parameters: $h = 1.0$, $g_\mathrm{inn} = 0.3$,  $g_\mathrm{out} = 0.5$, $\tilde g=0.9g_\mathrm{inn}$, $\kappa = \gamma^{z}_{i,i+1} = 0.2$, $\gamma^-_{i,i+1}=0$.}}
    \label{fig:SyncedMotifs}
\end{figure}

\textit{Synchronized Aharonov–Bohm Motifs.}---We now couple the $\mathscr C_n$-symmetric motifs introduced above and include collective dissipation to synchronize the spin dynamics across motifs. Each motif, labeled by $\nu$, is governed by its local Hamiltonian $H_n^\nu$; cf. Eq.~(\ref{eq:PlaquetteHamiltonian}). Inter-motif interactions are implemented via flip–flop couplings between corresponding spins in different motifs, described by 
\begin{equation}
H_I^{\nu \nu'} = \tilde g \sum_{\alpha \in \mathcal A}
\left(
\Plus{\alpha \nu} \Minus{\alpha \nu'} +
\Plus{\alpha \nu'} \Minus{\alpha \nu}
\right),
\end{equation}
where $\tilde g$ is the inter-motif coupling strength and $\mathcal A$ denotes all spin indices within a motif. Additionally, we include collective dissipation via $L_\nu = \sqrt{\kappa} \sum_{\alpha\in \mathcal A} \Z{\alpha\nu}$. 

The interplay of collective intra-motif dissipation and coherent inter-motif coupling leads to complete synchronization of the spin dynamics across all motifs, as the components of the dynamics that are not synchronized are damped out. To illustrate this mechanism, we analyze three coupled $\mathscr C_3$-symmetric motifs shown schematically in Fig.~\ref{fig:SyncedMotifs}a. The red outlines indicate collective dissipation acting on all spins in each motif, while the dashed lines denote the inter-motif couplings. The resulting magnetization dynamics for each motif is displayed in Figs.~\ref{fig:SyncedMotifs}b–d. 
After a transient time, the dynamics satisfies $\braket{\Z{\alpha\nu}(t)} =  \braket{\Z{\alpha\nu'}(t)}$ for $t\geq \tau$ for all motifs $\nu,\nu'$. The oscillating spins (brown and blue) synchronize fully in both amplitude and phase with their respective spins across the motifs, following the same pattern as in the single-motif case [cf. Fig.~\ref{fig:Motifs}], while the dissipative  {($\gamma^{z}_{i,i+1} \neq 0$,  $\gamma^{-}_{i,i+1} = 0$)} spins (pink) relax to identical stationary values across motifs. In contrast, without collective dissipation ($\kappa =0$), synchronization is lost completely as shown Figs.~\ref{fig:SyncedMotifs2} of Appendix B. Thus, it is the \emph{combined} action of local dissipation, correlated dissipation, and a $\pi$-flux that enables global synchronization across motifs. This mechanism is not limited to the $\mathscr C_3$ geometry: any $\mathscr C_n$-symmetric AB motif can be coupled in this way, allowing synchronized dynamics to emerge across arbitrary networks of AB motifs.

 {\textit{Discussion and Conclusions.}}---We have shown that the interplay of (synthetic) gauge fields and engineered dissipation can give rise to robust, synchronized spin dynamics in structures that we termed Aharonov–Bohm motifs. The local dissipation selectively damps components outside the interference-protected subspace, leaving behind oscillations that are stabilized by the motif’s rotational symmetry. We derived analytical expressions for the local magnetization and entanglement dynamics, confirmed by numerical simulations, and demonstrated how inter-motif coupling combined with correlated noise extends synchronization to larger networks.

 {These results establish a direct and previously unexplored connection between gauge fields, dissipation engineering, and quantum synchronization, and provide a robust mechanism for stabilizing collective many-body dynamics. More generally, our work identifies a regime in which quantum interference and dissipation act cooperatively: rather than destroying quantum coherences, dissipation stabilizes the interference-protected dynamics. An important implication of our results is their direct relevance for ongoing experimental efforts. Aharonov–Bohm caging has already been demonstrated in a variety of platforms, including photonic lattices and ultracold atoms~\cite{mukherjeeExperimentalObservationAharonovBohm2018, chenInteractiondrivenBreakdownAharonov2025b}. Our work therefore provides a concrete route towards realizing robust synchronized dynamics in these systems through engineered dissipation. More broadly, it suggests that interference-driven flat-band systems~\cite{vidalInteractionInducedDelocalization2000, creffieldCoherentControlInteracting2010a, mukherjeeObservationLocalizedFlatband2015, creutzEndStatesLadder1999a, zuritaTopologyInteractionsPhotonic2020a, vidalAharonovBohmCagesTwoDimensional1998} constitute a natural setting for exploring novel nonequilibrium quantum phenomena, where the interplay of gauge fields and  dissipation may give rise to new forms of collective behavior. 

Extending our framework to genuinely interacting many-body systems represents a particularly promising future direction. In this case, interactions may alleviate some of the fine-tuning requirements of the present setting, such as the need for an exact $\pi$-flux per plaquette or perfect rotational symmetry. Previous work \cite{Waechtler24} has shown that many-body interactions can replace fine-tuned (permutation) symmetries  and stabilize entire phases with synchronized observables. Many-body interactions in the closed system can destroy Aharonov-Bohm caging~\cite{liAharonovBohmCagingInverse2022c, chenInteractiondrivenBreakdownAharonov2025b}.  In the dissipative setting, however, the competition between interactions and engineered loss may instead stabilize new correlated phases or generate qualitatively different forms of synchronized many-body dynamics.  Furthermore, combining these ideas with many-body localized systems~\cite{danieliManybodyFlatbandLocalization2020b, danieliManybodyLocalizationTransition2022a} suggests an intriguing route towards stabilizing nonergodic yet synchronized phases, where dissipation enforces dynamical selection.}

\textit{Acknowledgements.}---The authors thank J. Liu and Q. Wu for helpful comments. G.P. acknowledges support from Spain’s MINECO through Grant No. PID2023-149072NBI00, CSIC Research Platform PTI-001, and National Project No. QTP2021-03-002. C.W.W. and G.P. received funding from the European Union’s Horizon Europe research and innovation programme under the Marie Skłodowska-Curie Actions (MSCA) grant agreement No. 101149948. Views and opinions expressed are, however, those of the author(s) only and do not necessarily reflect those of the European Union or the European Research Council. Neither the European Union nor the granting authority can be held responsible for them.

\bibliography{ref2,SM}

\onecolumngrid
\section*{End Matter}
\twocolumngrid

\begin{appendix}

\textit{Appendix A:  {Algebraic relations for} synchronization of local observables.}---Quantum synchronization, as discussed in the main text, refers to the emergence of persistent, synchronized oscillations of local observables in the long-time limit. Such oscillations are associated with purely imaginary eigenvalues of the Liouvillian $\mathcal L$. A robust route to guarantee the existence of these eigenmodes is  {given through specific algebraic relations}~\cite{Buca19,Buca22}: Consider a generic Lindblad master equation 
\begin{equation}
\label{eq:LindbladEndMatter}
    \dot\varrho = -\mathrm i \left[H, \varrho\right] + \sum\limits_\mu \left(L_\mu \varrho L_\mu^\dagger - \frac{1}{2}\left\{L_\mu^\dagger L_\mu, \varrho \right\}\right) \equiv \mathcal L[\varrho],
\end{equation}
with system Hamiltonian $H$ and Lindblad operators $L_\mu$.  {The system dynamics described by Eq.~(\ref{eq:LindbladEndMatter}) is guaranteed to have at least one steady state $\varrho_{\mathrm{ss}}$ such that $\mathcal L\left[\varrho_\mathrm{ss}\right] = 0$. A sufficient and necessary condition for the existence of a purely imaginary eigenvalue  $\lambda = -\mathrm i \omega$ of the Liouvillian with eigenstate  $\varrho = A \varrho_\mathrm{ss}$ of $\mathcal L$,  i.e., $\mathcal L[\varrho]=-\mathrm i \omega \varrho$ with $\omega \in \mathbb R$, is given by 
\begin{align}
\left[L_\mu ,A\right]\varrho_\mathrm{ss} &= 0, \label{eq:Theorem1a}\\
\left(-\mathrm i \left[H,A\right] - \sum\limits_\mu \left[L_\mu^\dagger, A\right]L_\mu \right)\varrho_\mathrm{ss} &= -\mathrm i \omega A \varrho_\mathrm{ss}.\label{eq:Theorem1b}
\end{align} 
}

While  {Eqs.~(\ref{eq:Theorem1a}) and (\ref{eq:Theorem1b}) guarantee} persistent oscillations in the steady-state manifold, synchronization of \emph{local} observables requires an additional permutation symmetry among subsystems. Let $P_{jk}$ denote the operator that exchanges subsystems $j$ and $k$ (e.g. spins in the AB motifs of the main text), and define the corresponding superoperator $\mathcal P_{jk}[x] = P_{jk} x P_{jk}$. If $\mathcal P_{jk}$ is a weak symmetry of the Liouvillian, $\left[\mathcal L, \mathcal P_{jk}\right] = 0$, and commutes with the operator $A$, i.e. $P_{jk} A P_{jk} = A$, then the dynamics of the two local observables $O_j$ and $O_k$ are synchronized whenever $\mathrm{Tr}\left[O_j \varrho_\pm\right] \neq 0$. That is, after a transient time $\tau$,
\begin{equation}
\braket{O_j(t)} = \braket{O_k(t)} \quad \forall~  t \geq \tau,
\end{equation}
up to exponentially small corrections. 

 {In the case of the main text where only spin-dephasing acts on the outer spins, i.e. $\gamma^-_{i,i+1}\equiv 0$, the algebraic relations can be simplified to the strong dynamical symmetry conditions: If there exists an operator $A$ that is an eigenoperator of the Hamiltonian, $\left[H, A\right] = \lambda A$, and simultaneously commutes with all Lindblad operators, $\left[L_\mu, A\right] = 0 = \left[L_\mu^\dagger, A\right]$, then the Liouvillain possesses at least one pair of purely imaginary eigenvalues $\pm \mathrm{i}\omega$, 
\begin{equation}
    \mathcal L [\varrho_\pm] = \pm \mathrm i\omega \varrho_\pm.
\end{equation}
}

\begin{figure}
    \centering
    \includegraphics[width=\linewidth]{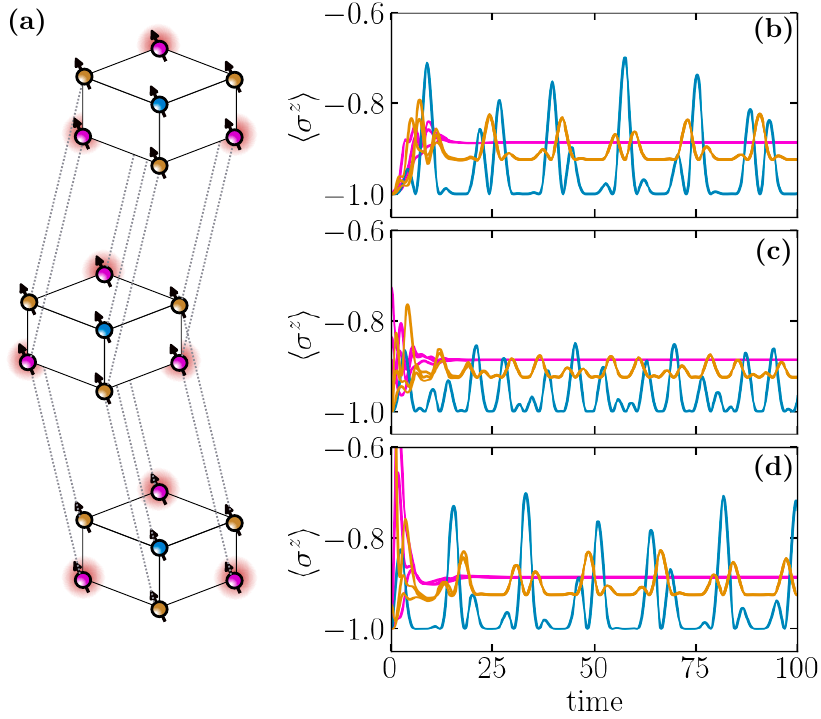}
    \caption{(a) Schematic representation of three coupled $\mathscr C_3$-symmetric motifs, which do not synchronize in the absence of collective dissipation (compared to Fig.~\ref{fig:SyncedMotifs}a in the main text) {, starting from the same random initial conditions~\cite{data}}. Panels (b)–(d) show the corresponding local magnetization dynamics for the motifs depicted to the left of each panel, now without the collective intra-motif dissipation. Although local dissipation is present, the long-term dynamics exhibits no synchronization, neither across motifs nor within each motif individually. Colors match those used in Fig.~\ref{fig:SyncedMotifs}(b–d) and the spin coloring in panel (a). 
 {Parameters: $h = 1.0$, $g_\mathrm{inn} = 0.3$,  $g_\mathrm{out} = 0.5$, $\tilde g=0.9 g_\mathrm{inn}$,  $\gamma^{z}_{i,i+1} = 0.2$, $\gamma^-_{i,i+1}=0$, $\kappa=0$.}}
    \label{fig:SyncedMotifs2}
\end{figure}

\textit{Appendix B: Loss of inter-motif synchronization.}---The emergence of fully synchronized spin dynamics across coupled AB motifs relies critically on the presence of additional collective intra-motif dissipation. To illustrate this explicitly, Figs.~\ref{fig:SyncedMotifs2}b--d show the local magnetization dynamics in the absence of collective dissipation (while retaining the same local loss processes), using identical parameters and initial conditions as in Fig.~\ref{fig:SyncedMotifs}. Each panel (b)--(d) displays the dynamics of the motif depicted to its left [panel (a)]. As discussed in the main text, when the collective channel is removed the synchronized oscillations disappear: not only is phase locking between motifs lost, but the inter-motif coupling also disrupts the residual intra-motif synchronization. 

\end{appendix}

\beginsupplement
\counterwithout{equation}{section}
\renewcommand{\theequation}{S\arabic{equation}}

\onecolumngrid
\begin{center}
\textbf{\large Supplementary Material to\\
`Synchronized Aharonov-Bohm Motifs via Engineered Dissipation'}
\end{center}
\setcounter{page}{1}

\section{Spectral decomposition of the Liouvillian superoperator}

Throughout this Supplementary Material, we frequently use a superoperator notation, which we summarize here for convenience. Consider a general Lindblad master equation of the form $\dot \varrho = \mathcal L [\varrho]$ where the Liouvillian superoperator $\mathcal L$ acts as
\begin{equation}
    \mathcal L[\varrho] = -\mathrm i \left[H, \varrho\right] + \sum\limits_\mu
\left(L_\mu \varrho L_\mu^\dagger - \frac{1}{2}\left\{L_\mu^\dagger L_\mu, \varrho\right\}\right).
\end{equation}
The Liouvillian can be written in its spectral decomposition,
\begin{equation}
\label{eq:SpectralDecomposition}
    \mathcal L  = \sum\limits_k \lambda_k \sket{\varrho_k}\sbra{\sigma_k},
\end{equation}
where $\lambda_k\in\mathbb C$ are the complex eigenvalues and $\sket{\varrho_k}$ and $\sbra{\sigma_k}$ denote the right and and left eigenstates of $\mathcal L$, respectively. These satisfy
\begin{equation}
    \mathcal L \sket{\varrho_k} = \lambda_k \sket{\varrho_k},\quad \mathcal L^\dagger \sket{\sigma_k} = \lambda^\ast_k \sket{\sigma_k}, \quad \sbraket{\sigma_k}{\varrho_{k'}} = \delta_{k,k'}, 
\end{equation}
with $\sbraket{\sigma}{\varrho} = \mathrm{Tr}(\sigma^\dagger \varrho)$  the Hilbert-Schmidt inner product. The formal solution of the master equation for an initial state $\varrho(0)$ is then
\begin{equation}
\label{eq:DynamicsDensityMAtrixSuper}
    \sket{\varrho(t)} = \sum\limits_k \exp(\lambda_k t)\sket{\varrho_k}\sbraket{\sigma_k}{\varrho(0)}.
\end{equation} 
Consequently, the expectation value of an observable $O$ evolves as
\begin{equation}
    \braket{O(t)} = \mathrm{Tr}\left[O\varrho(t)\right] = \sum\limits_k \exp(\lambda_k t)\sbraket{O}{\varrho_k}\sbraket{\sigma_k}{\varrho(0)}.
\end{equation}

\section{Synchronization condition for $\mathscr C_n$-symmetric Aharonov-Bohm motifs}

{In the following we will prove key properties, which must hold, for} the dissipative Aharonov-Bohm (AB) motifs with $\mathscr C_n$ rotational symmetry discussed in the main text to exhibit stable synchronization--meaning that generic initial states lead to synchronized dynamics of the local magnetization $\braket{\Z{i}(t)}$ after a transient time. {In particular, we will first consider the case where there is only $\sigma^{z}_{i,i+1}$ dissipators acting on the outer spins, in which case the total magnetization is a preserved quantity. For this case we will prove:}
\begin{enumerate}[label={(\roman*)}]
\item {Within} the single-magnon subspaces {of the Liouvillian} with total magnetization $m = \pm(N-2)$, there exists {a pair of purely imaginary eigenvalues which give} rise to the oscillatory Liouvillian eigenmodes responsible for synchronized dynamics of the local observables $\braket{\Z{i}(t)}$. 
\item All remaining magnetization sectors with $m \neq \pm(N-2)$ contain at most a single steady state and do not support oscillations; they contribute only a static offset and therefore do not affect synchronization.
\end{enumerate}
{It is then easy to verify that the synchronization subspace is also protected against local spin-relaxation $\sigma^-_{i,i+1}$ of the outer spins, which break  magnetization conservation and establish unidirectional coupling between sectors.}  

We first analyze a single four-spin plaquette and then generalize our arguments to arbitrary $\mathscr C_n$-symmetric motifs.

\subsubsection{Synchronization condition of a single spin plaquette with $N=4$}

We begin with a single plaquette of $N=4$ spins described by the Hamiltonian 
{\begin{equation}
H_{\mathrm{plaq}} = h\left(\Z{c} + \sum\limits_{i=1}^2 \Z{i} + \Z{1,2} \right) 
+  g_\mathrm{inn}  \sum\limits_{i=1}^2 \left(  e^{\mathrm i \phi_{i}} \Plus{c}\Minus{i} + \mathrm{h.c.}\right) + g_\mathrm{out} \left[e^{\mathrm i \phi_{1,2}}\left( \Plus{1}\Minus{1,2} + \Plus{1,2}\Minus{2}\right)
+ \mathrm{h.c.}\right],
\end{equation}}
where the total flux through the plaquette is $\phi = \sum_\square {\phi_\alpha}$ {with $\alpha$ either a single or double index}. {As mentioned previously, we first consider the case where only local spin-dephasing in terms of} the Lindblad operator {$L = \sqrt{\gamma^\mathrm{z}} \Z{1,2}$ acts on the plaquette}; cf. Eq.~(2). 

Because the plaquette is equivalent to a one-dimensional chain, the dynamics can be solved by mapping spin operators to spinless fermions through the Jordan–Wigner transformation. {We define our path along the plaquette as $c \to 1 \to (1,2) \to 2 \to c$. } The transformed Hamiltonian becomes
{\begin{equation}
\tilde H_{\mathrm{plaq}} = 2h\left(n_{c} + \sum\limits_{i=1}^2 n_{i} + n_{1,2} -2\right) 
+    \left[g_\mathrm{inn}  e^{\mathrm i \phi_{1}} c_c^\dagger c_1 + g_\mathrm{out} e^{\mathrm i \phi_{1,2}}\left( c^\dagger_{1}c_{1,2} + c^\dagger_{1,2}c_{2}\right) + \mathrm{h.c.}\right] + g_\mathrm{inn} (-1)^{(\bar n-1)} \left(e^{\mathrm i \phi_{2}} c_c^\dagger c_2 + \mathrm{h.c.}\right),
\end{equation}}
where {$\bar n = \sum_\alpha n_\alpha = \sum_\alpha c_\alpha^\dagger c_\alpha$} is the total fermion number. The fermionic and spin operators satisfy
{\begin{equation}
\Plus{\alpha} = \exp\left(-i\pi \sum_{\beta \prec \alpha} n_\beta\right) c_\alpha^\dagger, \quad
\Minus{\alpha} = \exp\left(i\pi \sum_{\beta \prec \alpha} n_\beta\right) c_\alpha
\end{equation}}
{where the sum is over all sites $\beta$ that come before $\alpha$ in the plaquette path.}

We first consider the single-excitation subspace (or single-magnon subspace in spin language).  {As only local spin-dephasing acts on the plaquette, synchronization} of local observables {is due to the existence of a} strong dynamical symmetry in the single-magnon sector, i.e., there exists an operator $A$ which is an eigenoperator of the Hamiltonian and that commutes with the dissipator $L = \sqrt{\gamma^z}(2n_{1,2} - 1) = L^\dagger$. For flux $\phi = \pi$, the Hamiltonian $H_\mathrm{plaq}$  {has eigenvalues  $E = \left\{-2h \pm \sqrt{2}g_\mathrm{inn}, -2h \pm \sqrt{2}g_\mathrm{out}\right\}$. The eigenstates associated with the energyeis $E_\pm = -2h \pm \sqrt{2}g_\mathrm{inn}$ given by 
\begin{equation}
    \ket{\psi_\pm} = \pm \frac{1}{\sqrt{2}} \ket{c} + \frac{1}{2} \left(\ket{1} +  \ket{2}\right).
\end{equation}
are eigenstates of $(2n_{1,2} - 1)$ (or equivalently $\Z{1,2}$ in the original spin picture)  with eigenvalue $-1$. Thus, the operator $A = \ket{\psi_-}\bra{\psi_+}$ fulfills the conditions for a strong dynamical symmetry (cf. Appendix A) as both states $\ket{\psi_\pm}$ have zero amplitude on the dissipative site. }
 The other eigenstates are not eigenstates of {$\Z{1,2}$} and thus do not satisfy the dynamical symmetry condition.

Within the resulting two-dimensional strong-dynamical symmetry subspace, the density matrix evolves as
{\begin{equation}
\varrho(t) = \braket{\psi_-|\varrho(0)|\psi_-} \varrho_{--} +  \braket{\psi_+|\varrho(0)|\psi_+} \varrho_{++} + \left[\mathrm{e}^{-\mathrm i \Omega t} \braket{\psi_+|\varrho(0)| \psi_-}\varrho_{+-} + \mathrm{h.c.}\right],
\label{eq:densitymatrixInDecFreeSubspace}
\end{equation}
where $\varrho_{\nu\nu'} = \ket{\psi_\nu}\bra{\psi_\nu'}$ with $\nu, \nu'=\pm$, and the oscillation frequency is  $\Omega = E_+ - E_- = 2\sqrt{2}g_{inn}$.} The local observables exhibit then synchronized oscillations; explicitly,
{\begin{equation}
    \braket{\Z c(t)} \propto -2 \left|\braket{\psi_+|\varrho(0)|\psi_-}\right|\cos(2\sqrt{2}g_\mathrm{inn} t) \qquad \braket{\Z 1(t)} = \braket{\Z 2(t)} \propto  \left|\braket{\psi_+|\varrho(0)|\psi_-}\right|\cos(2\sqrt{2}g_{\mathrm{inn}} t), 
\end{equation}}
demonstrating perfect synchronization among the non-dissipative spins.

Because both the Hamiltonian and dissipator are quadratic in the fermionic operators, every two-particle state is constructed from single-particle orbitals. For the double-excitation subspace to support nontrivial oscillatory dynamics, a two-dimensional dark subspace would be required; i.e., both fermions must occupy orbitals with zero amplitude on site {$\alpha=(1,2)$, asuring $n_{1,2} =0$} and thus have an eigenvalue of $-1$ of {$(2n_{1,2} - 1)$}. However, from the structure of the single-particle eigenstates, only one such two-particle state exists. Therefore the two-excitation sector contains only a \emph{one-dimensional} dark space and no strong dynamical symmetry. Consequently, it supports no oscillatory dynamics. The three-excitation (single-hole) subspace is equivalent to the single-excitation sector by particle–hole symmetry and therefore exhibits the same structure: one strong dynamical symmetry and the same oscillation frequency.

{Adding the additional disspator $L_- = \sqrt{\gamma^-}\sigma^-_{1,2}$ introduces unidirectional coupling between the different magnon subspaces. Furthermore, the strong dynamical symmetry is not preserved as $[L_-^\dagger, A]\neq 0$. However, the more general algebraic conditions summarized in Eqs.~(10) and (11) in Appendix A are still fulfilled. Thus, the synchronization subspace is still protected even if  the additional dissipative process breaks magnetization conservation and drives the population towards the fully de-excited state with $m = -4$. }

\subsubsection{Smallest motif with $\mathscr C_2$-rotational symmetry}

After establishing that the plaquette hosts a single strong dynamical symmetry in the one-magnon manifold, we now show that an analogous structure appears in the smallest AB motif with rotational symmetry. In contrast to the plaquette, no simple mapping to free fermions exists. Nevertheless, the 
$\mathscr C_2$ symmetry allows one to identify a two-dimensional {protected} subspace in the single-magnon sector, while all remaining magnetization sectors contain at most one dark state {if only local spin dephasing ($\gamma^-_{i,i+1}=0$) is acting on the system. In the case with additional local spin relaxation ($\gamma^-_{i,i+1}\neq 0$), there is again a cascaded dissipative process towards the fully de-excited state, while the synchronization subspace remains protected. }

The $\mathscr C_2$-symmetric motif consists of five spins; see the inset of Fig.~3a. As before, we seek eigenstates with zero weight on the two outer spins where dissipation acts. Because of the rotation symmetry, this condition must hold for both outer spins simultaneously, since $C_2\Z{1,2} C_2^{-1} = \Z{2,1}$ and vice versa.

The Hamiltonian is given by $H_n$ of the main text with $n=2$ [see Eq.~(1)], and the dissipative dynamics by the Lindblad equation in Eq.~(2). Projecting into the one-magnon subspace using 
 $\mathcal P_n = \sum\limits_{\alpha} \ket{\alpha}\bra{\alpha}$, 
 with $\ket{\alpha}$ denoting a configuration with a single spin-up excitation, we obtain the projected Hamiltonian for the $\mathscr C_2$-symmetric motif $\tilde H_2 =\mathcal PH_2 \mathcal P$. With the gauged fixed such that the flux per plaquette is $\phi = \pi$ and at the outer boundaries of the plaquettes, 
{\begin{equation}
\tilde H_2 =
-2h \sum_\alpha \ket{\alpha}\bra{\alpha}
+ g_\mathrm{inn}\big(
\ket{c}\bra{1}
+ \ket{c}\bra{2}+ \mathrm{h.c.}\big)
+ g_\mathrm{out}\big(
\mathrm{i}\ket{(1,2)}\bra{1}
+ \mathrm{i}\ket{2}\bra{(1,2)}
+ \mathrm{i}\ket{(2,1)}\bra{2}
+ \mathrm{i}\ket{1}\bra{(2,1)}
+ \mathrm{h.c.}
\big).
\end{equation}}
A state with no excitation on the outer sites $\alpha = (1,2)$ or $\alpha = (2,1)$ takes the form  ${\ket{\phi}} = a_c \ket{c} + a_1\ket{1} + a_2 \ket{2}$. Since, {$[\tilde H_2, C_2] = 0$}, eigenstates of {$\tilde H_2$} can be chosen to satisfy {$C_2 \ket{\phi} = \pm \ket{\phi}$}. This yields the symmetry-adapted combinations 
{\begin{equation}
\ket{\phi_\mathrm{sym}} = a_c \ket{c} + a_1 \left(\ket{1} +  \ket{2}\right) \qquad \ket{\phi_\mathrm{anti}} = a_c \ket{c} + a_1 \left(\ket{1} -  \ket{2}\right).
\end{equation}}
 The antisymmetric combination {$\ket{\phi_\mathrm{anti}}$} is not an eigenstate of $\tilde H_2$, while the symmetric state {$\ket{\phi_\mathrm{sym}}$} satisfies
{\begin{equation}
\tilde H_2 \ket{\phi_\mathrm{sym}}
= -2h \ket{\psi_+}
+ a_c g_\mathrm{inn} (\ket{1} + \ket{2}) + 2 g_\mathrm{inn} a_1 \ket{1}
= E \ket{\psi_+}.
\end{equation}}
This yields $a_c = \pm \sqrt{2} a_1$, corresponding to two eigenstates with energies {$E_\pm =-2h \pm \sqrt{2}g_\mathrm{inn}$}. 
Fixing normalization sets $a_1 = 1/2$. Thus, the single-magnon subspace contains a two-dimensional {protected} subspace spanned by two eigenstates with distinct energies. Consequently, the resulting dynamics of the density matrix is analogous to Eq.~(\ref{eq:densitymatrixInDecFreeSubspace}).

We now examine the two-excitation manifold. A dark state must be an eigenstate of both dissipators $\Z{1,2}$, $\Z{2,1}$, and the rotation operator $C_2$. Let {$\ket{\chi}$} satisfy
{\begin{equation}
\Z{1,2}\ket{\chi} = z_{1,2} \ket{\chi},\qquad  \Z{2,1}\ket{\chi} = z_{2,1} \ket{\chi},\qquad  C_2 \ket{\chi} = p \ket{\chi},
\end{equation}}
where $z_{1,2}, z_{2,1}, p \in \{\pm 1\}$. Using $C_2 \Z{1,2} = \Z{2,1}C_2$ gives
 {\begin{equation}
      z_{1,2} p \ket{\chi} = \Z{1,2} C_2 \ket{\chi} = C_2 \Z{2,1} \ket{\chi} = z_{2,1} p \ket{\chi},
 \end{equation}}
which implies $z_{1,2} = z_{2,1}$. If both equal $+1$, the excitations occupy only the outer spins; this configuration is not an eigenstate of the Hamiltonian because hopping is allowed on the connected graph. The only remaining case is $z_{1,2} = z_{2,1} = -1$, implying that the state has no support on the outer spins.

Restricting to the corresponding inner-excitation subspace,
 $\mathcal P_\mathrm{inner} = \mathrm{span}\{\ket{c1},\ket{c2}, \ket{12}\}$, the action of $C_2$ gives $C_2 \ket{c1} = \ket{c2}$ and $C_2 \ket{12} = \ket{12}$. Thus, the subspace decomposes into symmetric and antisymmetric sectors according to the eigenvalue of $C_2$:
\begin{equation}
\mathcal P_\mathrm{inner}^{(+)} = \mathrm{span}\left\{ \ket{12}, \tfrac{1}{\sqrt{2}}(\ket{c1} + \ket{c2}) \right\},
\qquad
\mathcal P_\mathrm{inner}^{(-)} = \mathrm{span}\left\{ \tfrac{1}{\sqrt{2}}(\ket{c1} - \ket{c2}) \right\}.
\end{equation}
 We now examine whether these subspaces are invariant under the Hamiltonian $\tilde H_2$. Acting with $\tilde H_2$ on states of $\mathcal P_\mathrm{inner}^{(+)}$ shows that they are not preserved within $\mathcal P_\mathrm{inner}^{(+)}$, that is, they `leak' into states involving the outer spins. In contrast, the antisymmetric subspace $\mathcal P_\mathrm{inner}^{(-)}$ remains invariant under $\tilde H_2$. However, as $\mathcal P_\mathrm{inner}^{(-)}$ is one-dimensional, it contains only a single such state and therefore does not support a two-dimensional subspace capable of sustaining oscillatory dynamics.
Thus, the two-magnon manifold contributes only static, non-oscillatory corrections to long-time local observables.
 
\subsubsection{General motif with $\mathscr C_n$-symmetry}

We now extend the previous analysis to a motif composed of $n$ plaquettes with $\mathscr C_n$-rotational symmetry. Using the same symmetry constraints as before, we show that synchronized oscillations arise exclusively within the one-magnon manifold, while all other magnon sectors contribute only static long-time dynamics.

Dark states must be simultaneous eigenstates of the rotation operator $C_n$ and of all $\Z{i,i+1}$. Since $C_n \Z{i,i+1} C_n^{-1} = \Z{i+1,i+2}$, the eigenvalue conditions enforce $z_{i,i+1} = -1$ for all $i$, implying that a dark state must have zero weight on all outer spins. Solving these constraints yields precisely two one-magnon states, 
\begin{equation}
\label{appEq:PsiPlusMInus}
\ket{\psi_\pm} = \frac{1}{\sqrt{2}}\left(\pm \ket{c} + \frac{1 }{\sqrt{n}} \sum_{i=1}^n \ket{i}\right),
\end{equation}
with energies {$E_\pm = -2h \pm \sqrt{n} g_\mathrm{inn}$}. Thus, each single-magnon sector contains a two-dimensional {protected synchronization} subspace spanned by $\ket{\psi_\pm}$.

For two excitations, the `no-leakage' condition (vanishing amplitude on all outer spins) translates to
\begin{equation}
\sigma_{i,i+1}^+ \big(\sigma_i^- - \sigma_{i+1}^-\big) \ket{\psi} = 0 \quad \forall i.
\end{equation}
Eploiting $\mathscr C_n$ symmetry, a convenient basis for the inner two-excitation subspace is
\begin{equation}
\ket{C_l} = \frac{1}{\sqrt{n}} \sum_i \lambda^{l i} \ket{c i},
\qquad
\ket{I_l} = \frac{1}{\sqrt{x}} \sum_i \lambda^{l i} \ket{ii+1},
\qquad
\lambda = e^{2\pi \mathrm i / n}.
\end{equation}
Evaluating the constraint shows that only the totally symmetric state $\ket{C_0}$ satisfies the no-leakage condition; all other states necessarily hybridize with outer-spin configurations and are therefore not dark. Hence, the two-magnon manifold contains a single dark state.

The same argument extends to all higher-excitation sectors: only the one-magnon spaces possess a two-dimensional invariant subspace capable of supporting coherent oscillations. All higher-magnon sectors contribute solely static, non-oscillatory corrections to the long-time behavior of local observables. {As mentioned previously, it is easy to check that the subspcae fulfills the algebraic constraints of Eqs.~(10) and (11) of Appendix A even if additional local spin relaxation, i.e, $\gamma^-_{i,i+1}\neq0$, acts on the system. Thus, also in the case of local spin relaxation the synchronized dynamics is preserved.}

\section{Analytic expression of the local spin magnetization dynamics}

As established above, only the single-magnon manifolds contribute to oscillatory dynamics; all higher-excitation sectors yield static long-time behavior. We therefore restrict the analysis to the one-magnon subspace. Each such subspace hosts a single strong dynamical symmetry and an associated two-dimensional manifold spanned by the eigenstates $\ket{\psi_\pm}$; cf. Eq.~(\ref{appEq:PsiPlusMInus}). In all examples considered in the main text, we find precisely one additional stationary mode  $\bar \varrho$ not connected to the dynamical-symmetry sector, satisfying $\mathcal L[\bar \varrho] = 0$. For clarity, we include only this mode; additional stationary contributions from other excitation sectors can be evaluated analogously.

Using the spectral decomposition of the Liouvillian, Eq.~(\ref{eq:SpectralDecomposition}), the long-time limit of the density matrix from Eq.~(\ref{eq:DynamicsDensityMAtrixSuper}) becomes
\begin{equation}
\lim\limits_{t\to\infty}\sket{\varrho(t)} = \sket{\bar \varrho}\sbra{\bar \sigma} + \sket{\varrho_{++}}\sbra{\sigma_{++}} + \sket{\varrho_{--}}\sbra{\sigma_{--}} + e^{\mathrm i \Omega t}  \sket{\varrho_{+-}}\sbra{\sigma_{+-}} + e^{-\mathrm i \Omega t} \sket{\varrho_{-+}}\sbra{\sigma_{-+}},
\end{equation}
where $\sket{\varrho_{\nu\nu'}}$ denotes the vectorized operator $\ket{\psi_\nu}\bra{\psi_{\nu'}}$ with $\ket{\psi_\nu}$ given by Eq.~(\ref{appEq:PsiPlusMInus}) for $\nu=\pm$. 

The expectation value of any operator $\Z{\alpha}$ follows as 
\begin{equation}
\label{appEq:MagnetizationExpression}
    \lim\limits_{t\to\infty}\braket{\Z{\alpha}(t)} = \sbraket{\Z{\alpha}}{\bar \varrho}\sbraket{\bar \sigma}{\varrho(0)} + \sum\limits_{\nu=\pm}\sbraket{\Z{\alpha}}{ \varrho_{\nu\nu}}\sbraket{\sigma_{\nu\nu}}{\varrho(0)} + e^{\mathrm i \Omega t}\sbraket{\Z{\alpha}}{\varrho_{+-}}\sbraket{\sigma_{+-}}{\varrho(0)} + e^{-\mathrm i \Omega t}\sbraket{\Z{\alpha}}{ \varrho_{-+}}\sbraket{\sigma_{-+}}{\varrho(0)}.
\end{equation}
Note, that $\sbraket{\sigma_{\nu\nu'}}{\varrho(0)} = \braket{\psi_\nu|\varrho(0)|\psi_{\nu'}}$ and similarly $\sbraket{\Z{\alpha}}{ \varrho_{\nu\nu'}} = \braket{\psi_{\nu'}|\Z{\alpha}|\psi_{\nu}}$. Evaluating the different terms of Eq.~(\ref{appEq:MagnetizationExpression}) by plugging in Eq.~(3), one then finds the expressions of the main text for the local spin magnetization, namely
\begin{equation}
    \braket{\Z{c}(t)} =  A_c - 2 \left|\braket{\psi_+|\varrho(0)|\psi_-}\right|\cos(\Omega t+\varphi),\quad \braket{\Z{i}(t)} = \bar A + \frac{2}{n} \left|\braket{\psi_+|\varrho(0)|\psi_-}\right|\cos(\Omega t+\varphi), \quad \braket{\Z{i,i+1}(t)} = \tilde A.
\end{equation}
The constants are given by 
\begin{align}
     A_c &= \sbraket{\Z{c}}{\bar \varrho}\sbraket{\bar \sigma}{\varrho(0)},\\
     \bar A &= \sbraket{\Z{i}}{\bar \varrho}\sbraket{\bar \sigma}{\varrho(0)} + \left(\frac{1}{n}-1\right)\left(\braket{\psi_+|\varrho(0)|\varrho_+} + \braket{\psi_-|\varrho(0)|\varrho_-}\right),\\
     \tilde A &= \sbraket{\Z{i,i+1}}{\bar \varrho}\sbraket{\bar \sigma}{\varrho(0)} - \left(\braket{\psi_+|\varrho(0)|\varrho_+} + \braket{\psi_-|\varrho(0)|\varrho_-}\right)
\end{align}
Note that $\braket{\psi_\nu|\Z{c}|\psi_\nu} = 0$ and $\braket{\psi_\nu|\Z{i}|\psi_\nu} = -1 + 1/n$.

\section{Analytic expression of concurrence between two spins}

To characterize the entanglement generated by synchronized dynamics in a $\mathscr C_n$-symmetric motif, we evaluate the concurrence $\mathcal C$, defined by
\begin{equation}
\mathcal C(\varrho_{\alpha\alpha'}) = \max\{0, \lambda_1 - \lambda_2 - \lambda_3 - \lambda_4\},
\end{equation}
where $\varrho_{\alpha\alpha'} = \mathrm{Tr}_{\mathcal M_n \setminus (\alpha,\alpha')}[\varrho(t)]$ is the reduced density matrix obtained by tracing out all spins of the motif except $\alpha$ and $\alpha'$. The quantities $\lambda_k$ denote the square roots of the ordered eigenvalues of the matrix $\varrho_{\alpha\alpha'} \tilde{\varrho}_{\alpha\alpha'}$, where $\tilde{\varrho}_{\alpha\alpha'}$ is the spin-flipped state \cite{woottersEntanglementFormationArbitrary1998}.

Within the single-excitation subspace, the reduced density matrix takes the form
\begin{equation}
    \varrho_{\alpha \alpha'} = \left(
\begin{matrix}
    p_{00} & 0 & 0 & 0 \\
    0 & p_{10} & \varrho_{10,01} & 0 \\
    0 &  \varrho_{01,10} & p_{01} &  0 \\
    0 & 0 & 0 & 0 \\
\end{matrix}
    \right),
\end{equation}
where $p_{00}$ is the probability of no excitation in either spin $\alpha$ and $\alpha'$, $p_{10}$  and $p_{01}$ the probability of the excitation occupying $\alpha$ or $\alpha'$, respectively, and $\varrho_{10,01}$  the corresponding coherence. This is an X-state for which the concurrence reduces to \cite{zhaoCoherenceConcurrenceStates2020}
\begin{equation}
     \mathcal C(\varrho_{\alpha\alpha'}) = 2 \left| \varrho^{10,01}_{\alpha \alpha'}\right| = 2 \left| \braket{\Plus{\alpha}\Minus{\alpha'}}\right|.
\end{equation}

The time dependence follows from the spectral decomposition of the Liouvillian, Eqs.~(\ref{eq:SpectralDecomposition}) and (\ref{eq:DynamicsDensityMAtrixSuper}). We isolate the contribution from the protected subspace spanned by $\ket{\psi_\pm}$, while the additional steady-state contribution $\tilde \varrho$ can be obtained numerically. Owing to $\mathscr C_n$ symmetry, it suffices to evaluate the concurrence either between the central spin and an inner spin, or between two inner spins. 

Using the definitions of $\ket{\psi_\pm}$ in Eq.~(3) we find
\begin{equation}
    \mathcal C(\varrho_{ci}) = 2 \left| \braket{\ket{c}\bra{i}}\right| = \frac{1}{\sqrt{n}}\left| \braket{\psi_+|\varrho(0)|\psi_+} - \braket{\psi_-|\varrho(0)|\psi_-} - 2 \mathrm{i}\left|\braket{\psi_+|\varrho(0)|\psi_-}\right| \sin(\Omega t + \varphi)\right|, 
\end{equation}
where $\braket{\psi_+|\varrho(0)|\psi_-}= |\braket{\psi_+|\varrho(0)|\psi_-}| e^{\mathrm{i}\varphi}$.
In general, $\tilde{\varrho}$ yields an additional (static) contribution.  

Similarly,
\begin{equation}
        \mathcal C(\varrho_{ij}) = 2 \left| \braket{\ket{i}\bra{j}}\right| = \frac{1}{{n}}\left| \braket{\psi_+|\varrho(0)|\psi_+} + \braket{\psi_-|\varrho(0)|\psi_-} +2 \left|\braket{\psi_+|\varrho(0)|\psi_-}\right| \cos(\Omega t + \varphi)\right|. 
\end{equation}

For the special case that the initial state is a pure state of the form $\varrho(0) = \ket{\psi_0}\bra{\psi_0}$ with
\begin{equation}
    \ket{\psi_0} = \frac{1}{\sqrt{1+b^2}}\left(\ket{\psi_+} + b \ket{\psi_-} \right),
\end{equation}
the concurrences are given by
\begin{align}
 \mathcal C(\varrho_{ci}) &= \frac{1}{(1+b^2)\sqrt{n}}\sqrt{\left(1-b^2\right) + 4 b^2 \left[\sin(\Omega t)\right]^2}, \\   
 \mathcal C(\varrho_{ij}) &= \frac{1}{2n} + \frac{b}{(1+b^2)n}\cos(\Omega t).
\end{align}
For $b = 1$, i.e., an equal superposition of the two dynamical-symmetry eigenstates, the central–outer concurrence $\mathcal C(\varrho_{ci})$ oscillates between $0$ and $1/\sqrt{n}$, while outer–outer concurrence $\mathcal C(\varrho_{ij})$ oscillates between  $0$ and $1/n$. The oscillation amplitudes therefore decrease with increasing $n$, reflecting the delocalized structure imposed by the $\mathscr C_n$ symmetry of the system. Moreover, even when the oscillatory contribution vanishes, the static contribution originating from the steady state remains bounded by the same values, since rotational symmetry enforces an equal distribution of the single excitation over the inner spins.

{To understand why the entanglement can vanish at certain times even though the spins remain correlated, let us look at the $\mathscr C_2$-symmetric motif, and trace out the central spin and examine the reduced dynamics of the two inner spins explicitly. The states forming the synchronization subspace are 
\begin{equation}
\ket{\psi_\pm} = \pm a_c \ket{c} + a_1 \left(\ket{1} +  \ket{2}\right),
\end{equation}
such that the dynamics is given by oscillations between these two states, in particular 
\begin{equation}
    \ket{\psi(t)} = (\ket{\psi_+} e^{\mathrm{i} \Omega t} + \ket{\psi_-} e^{-\mathrm{i} \Omega t})/\sqrt{2} = 2\mathrm i a_c \sin (\Omega t) \ket{c} + 2 a_1 \cos(\Omega t ) (\ket{1} + \ket{2}),
\end{equation}
where the excitation oscillates between the central spin and the two inner spins. Tracing out the central spin yields:
\begin{equation}
    \rho_{12}(t) = 4a_c^2\sin^2(\Omega t)\ket{\downarrow\downarrow}\bra{\downarrow\downarrow} + 4a_1^2\cos^2(\Omega t)\big(\ket{\uparrow\downarrow}+\ket{\downarrow\uparrow}\big)\big(\bra{\uparrow\downarrow}+\bra{\downarrow\uparrow}\big).
\end{equation}
The physical picture is then transparent: when the excitation resides on the central spin ($\cos(\Omega t) = 0$), the two inner spins are in the product state $\ket{\downarrow\downarrow}$ and carry no entanglement. Conversely, when the excitation resides on the inner spins ($\sin(\Omega t)=0$), they form a Bell-like state $\ket{\uparrow\downarrow}+\ket{\downarrow\uparrow}$ and are maximally entangled. The entanglement between the inner spins therefore oscillates periodically, vanishing whenever the excitation is transferred to the central spin. Crucially, however, the spins remain correlated throughout the dynamics, i.e., the vanishing of entanglement at certain times does not signal an absence of quantum correlations, but rather reflects the redistribution of the excitation across the motif. This confirms that the synchronized dynamics is a genuinely quantum phenomenon, rather than merely classically correlated spin flips. }

\section{Liouvillian Perturbation theory}

We derive here the second-order correction to the purely imaginary Liouvillian eigenvalues associated with synchronized dynamics under coherent perturbations of the Hamiltonian. As discussed in the main text, perturbations of the dissipative rates do not alter the qualitative structure of the dynamics. We therefore consider a perturbed Hamiltonian $H + \varepsilon V$, leading to a perturbed Liouvillian $\mathcal L = \mathcal L_0 +\varepsilon \mathcal V$, where $\mathcal L_0$ is the Liouvillian superoperator of the unperturbed systems and $\mathcal V[\varrho] = -\mathrm i \left[V, \varrho\right]$. We focus on the complex-conjugate pair of purely imaginary eigenvalue $\lambda_\pm$ with eigenstates $\sket{\varrho_\pm}$. Since they are always a complex conjugate pair (even after perturbation due to the structure of $\mathcal L$), we specifically focus on $\lambda_+$. 

The first order equation of the (non-degenerate) perturbation theory is given by 
\begin{equation}
\label{appEq:FirstOrder}
    \mathcal L_0 \kket{\varrho_+^{(1)}} + \mathcal V \kket{\varrho_+^{(0)}} = \lambda_+ \kket{\varrho_+^{(1)}} + \lambda_+^{(1)} \kket{\varrho_+^{(0)}},
\end{equation}
where $\kket{\varrho_+^{(0)}} = \kket{\varrho_+}$ of the unperturbed system. Imposing $\sbraket{\sigma_+^{(0)}}{\varrho_+^{(1)}} = 0$ leads to the first-order correction 
\begin{equation}
    \lambda_+^{(1)} = \kbraket{\sigma_+^{(0)}}{\mathcal V}{\varrho_+^{(0)}}.
\end{equation}
As $\mathcal V = -\mathrm i \left[V, \cdot\right]$, the matrix element is purely imaginary. Thus, the first-order correction yields only a frequency renormalization and does not introduce decay. We therefore proceed to the second-order term.

The second order correction to the eigenvalue is given by
\begin{equation}
    \lambda_+^{(2)} = \kbraket{\sigma_+^{(0)}}{\mathcal V}{\varrho_+^{(1)}}.
\end{equation}
To evaluate the second-order, we need to determine the first-order correction to the eigenstate. In general we can express it as 
\begin{equation}
    \kket{\varrho_+^{(1)}} =  \sum\limits_{l\neq +}c_l \kket{\varrho_l^{(0)}}.
\end{equation}
Then from the first order Eq.~(\ref{appEq:FirstOrder}) we find
\begin{equation}
    c_l = \frac{\kbraket{\sigma_l^{(0)}}{\mathcal V}{\varrho_+^{(0)}}}{\lambda_+ - \lambda_l}. 
\end{equation}
The second order correction to the eigenvalue is then 
\begin{equation}
    \lambda_+^{(2)} = \sum\limits_{l\neq+} \frac{1}{\lambda_+ - \lambda_l}\kbraket{\sigma_+^{(0)}}{\mathcal V}{\varrho_l^{(0)}}\kbraket{\sigma_l^{(0)}}{\mathcal V}{\varrho_+^{(0)}}.
\end{equation}
Because $\lambda_+$ is purely imaginary while all other $\lambda_l$ have strictly negative real parts, the denominator generally acquires a non-zero real component, implying that the second-order correction $\lambda_+^{(2)}$ generically acquires a real part of order $\varepsilon^2$. 

Thus, Hamiltonian perturbations of strength $\varepsilon$ induce decay only at order $\mathcal O(\varepsilon^2)$. As long as the perturbation remains small, the imaginary part of $\lambda_+$, and hence the synchronized oscillations, remains robust and observable over long times.


\end{document}